\documentclass[conference]{IEEEtran}
\IEEEoverridecommandlockouts

\ifCLASSINFOpdf
   \usepackage[pdftex]{graphicx}
\else
\fi

\usepackage{multirow}
\usepackage{array}
\usepackage[lofdepth,lotdepth]{subfig}

\begin{document}

%\IEEEpubid{0000--0000/00\$00.00˜\copyright˜2015 IEEE
%}
%\IEEEpubid{\makebox[\columnwidth]{ 978-1-7281-8073-1/20/~\$31.00
%\copyright ~ 2020 IEEE\hfill} \hspace{\columnsep}\makebox[\columnwidth]{}}

\title{Detection of Covid-19 Patients with Convolutional Neural Network Based Features on Multi-class X-ray Chest Images}

\author{\IEEEauthorblockN{Ali Narin$^{1}$}
\IEEEauthorblockA{$^{1}$Department of Electrical and Electronics Engineering, Zonguldak Bulent Ecevit University, Zonguldak, Turkey\\
{alinarin@beun.edu.tr}\\
}
}


%Telephone: (800) 555--1212, Fax: (888) 555--1212}


\maketitle

\begin{abstract}
Covid-19 is a very serious deadly disease that has been announced as a pandemic by the world health organization (WHO). The whole world is working with all its might to end Covid-19 pandemic, which puts countries in serious health and economic problems, as soon as possible. The most important of these is to correctly identify those who get the Covid-19. Methods and approaches to support the reverse transcription polymerase chain reaction (RT-PCR) test have begun to take place in the literature. In this study, chest X-ray images, which can be accessed easily and quickly, were used because the covid-19 attacked the respiratory systems. Classification performances with support vector machines have been obtained by using the features extracted with residual networks (ResNet-50), one of the convolutional neural network models, from these images. While Covid-19 detection is obtained with support vector machines (SVM)-quadratic with the highest sensitivity value of 96.35\% with the 5-fold cross-validation method, the highest overall performance value has been detected with both SVM-quadratic and SVM-cubic above 99\%. According to these high results, it is thought that this method, which has been studied, will help radiology specialists and reduce the rate of false detection. \\
\end{abstract}


% \fi
%\IEEEpeerreviewmaketitle

%\IEEEpubidadjcol

\section{Introduction}
Coronavirus is an infectious virus seen in animals and humans. The new coronavirus disease, a member of the coronavirus family, which broke out in late December is caused by the SAR-CoV-2 virus \cite{Xu}. This infectious disease, known as Covid-19 in the literature, attacks the respiratory system directly, so symptoms of fever, cough, and shortness of breath are frequently observed. In advanced cases, it causes inflammation of the air sacs in the lungs such as known as pneumonia \cite{Roudsari}.

The epidemic's spread around the world has seriously damaged the economies and health systems of countries. For this reason, the biggest effort of all countries recently is to end the epidemic. For this, it is very important to detect and isolate those who have Covid-19 disease at an early stage. Detection of the presence of Covid-19 by reverse transcription polymerase chain reaction (RT-PCR) tests is of vital importance \cite{Waller}. Due to its low sensitivity and time consuming, alternative fast and effective methods are needed to be developed. In this context, there are some studies conducted with lung imaging methods (X-ray or CT). Some of them are as follows: Studies with feature extraction and classification \cite{Ozturk,Wei,Elaziz}, studies performed with convolutional neural networks without external feature extraction, which are among the end-to-end methods \cite{Narin,Mahmud,Talo} are studies with segmentation methods \cite{Fan,Wu}.

In this study, a 3-class (Covid-19, Normal, Viral Pneumonia) detection study has been carried out with the help of SVM using feature maps obtained from the ResNet-50 model, which is a convolutional neural network (CNN). 1x1000 attribute map was obtained from each image. The study has been carried out in 10 replicates with 5-fold cross validation method.

In the following sections, the data used in the study, the convolutional neural network and model used, support vector machines, performances criteria and results will be mentioned.
\\ 


%Place this somewhere in the second column of the first page :\IEEEpubidadjcol
\IEEEpubidadjcol

\subsection{Database}
In this study, 3-class datasets are used, which are available to everyone on the Kaggle website. The data set consists of 219 "Covid-19", 1341 "Normal" and 1345 "Viral Pneumonia" X-ray chest images shown as an example in Fig 1 \cite{Rahman}. Each image in the data set is 1024x1024 data.

\begin{figure}[ph]
\centering
\includegraphics[width=3.5in]{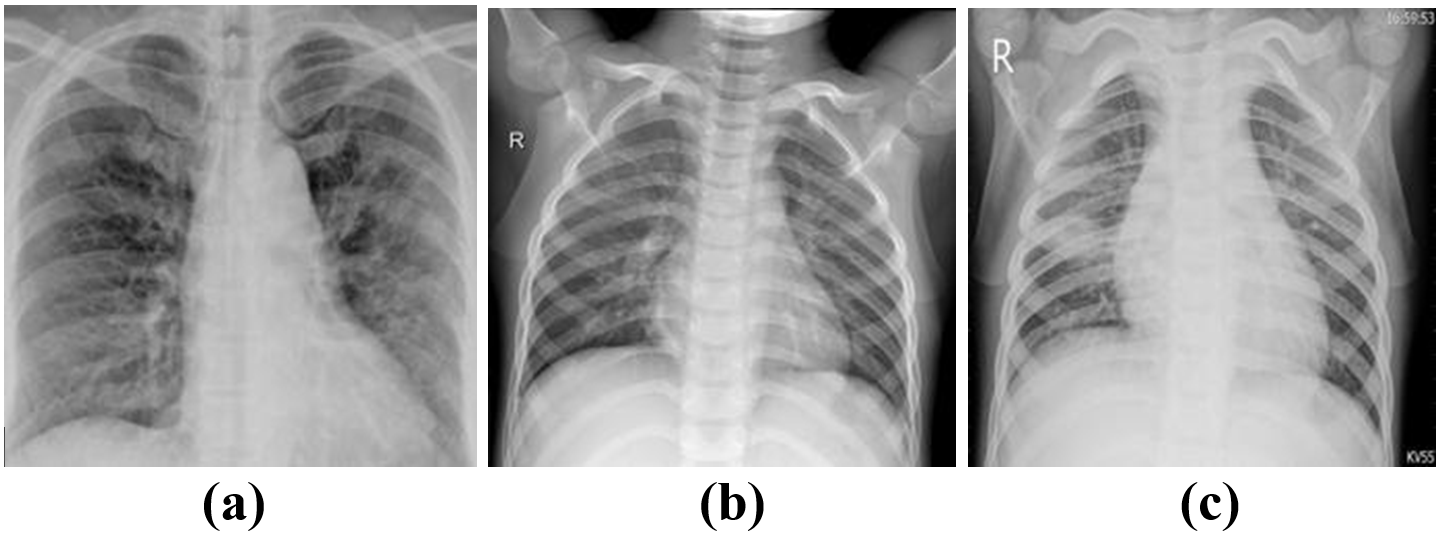}
% where an .eps filename suffix will be assumed under latex, 
% and a .pdf suffix will be assumed for pdflatex; or what has been declared
% via \DeclareGraphicsExtensions.
\caption{X-ray images a) Covid-19 b) Normal c) Viral Pneumonia.}
\label{fig_sim}
\end{figure}

\subsection{Convolutional Neural Networks}
The most widely known and used deep learning models are convolutional neural networks. CNN consists of one or more convolution layers, pooling layers and fully connected layers. Dropout etc. can be added in different layers. One of the biggest advantages of CNN models is that the model is realized using data directly in the training process \cite{Schmid}. The extraction of feature maps takes place in the convolution and pooling layers. The classification process takes place at the end of the fully connected layer.

Defining Data: Standard size data is given as input to the network. The important thing here is the size of the data. Performance may vary depending on the model used.

Convolution layers: The convolution layer is the most basic layer from which CNN is named. By sliding the filters used here over the image, the feature maps of the image are created at the output.

Pooling layers: In this layer, the prominent parts of the obtained feature maps are obtained. The feature map is updated with the smallest, largest, average values etc. among the measures determined on the map. However, there may be loss of information in this layer.. 

Fully Connected Layers: It is the layer added to the model after the feature maps are created. It has a feature that brings together all the layers before it. For example, 10x10x10 data is converted to 1x1000 and operations are performed.

\subsection{Feature Extraction using ResNet-50 Convolutional Neural Network}

The use of CNN models has become quite common in recent years. There are many models in the literature. The fact that the models have different depths and the number of data increases day by day significantly affect the training period. For this, the use of pre-trained models is preferred to overcome this problem. In this study, the ResNet-50 model consisting of 5 convolution blocks and 50 layers in total was used \cite{He}. 1x1, 3x3 and 1x1 operations are applied in convolution blocks. Different feature maps are obtained from images with filters of different sizes. The features used in the study have been obtained from the fully connected layer in 1000x1 dimensions.

\subsection{Support Vector Machines}
SVM algorithm, which was introduced for the first time in 1963, is a supervised learning approach based on the statistical learning theory \cite{Vapnik,Vapnik2}. It is widely used in linear and nonlinear classification problems. The purpose of the algorithm is to determine the support vectors that provide the clearest distinction between classes and to draw the line called the hyper plane. It provides separation of classes that are not separated linearly from kernel functions. In this study, results for different functions have been obtained by using linear, cubic and quadratic functions.

\subsection{Performance Metrics}
The classification performance of the obtained features on SVM was measured with 3 parameters. These are expressed as \cite{NarinTip}:
\begin{eqnarray}
	SEN&=&\frac{TP}{TP+FN}\\
	SPE&=&\frac{TN}{TN+FP}\\
	ACC&=&\frac{TP+TN}{TP+TN+FP+FN}
\end{eqnarray}
Here, the number of those who are really Covid-19 patients and found as Covid-19 patients by the classifier is expressed as TP and the number of those who are mistakenly found in the form of other classes as FN, the number of those who are really non-Covid-19 patients and found as non-Covid-19 patients by the classifier is expressed as TN and the number of those who are mistakenly found in Covid-19 patients by the classifier is expressed as FP. Using these values, the classifier performances, sensitivity (SEN), specificity (SPE) and general performance (ACC) values were calculated. The SEN value gives the performance of those with Covid-19, the performance of the classes whose SPE value is out of Covid-19, and the ACC is general performance values.

5-fold cross validation method was used to obtain a more stable result. As shown in Fig 2, the whole data set is divided into 5-fold.  One part of it is reserved for testing, while (k-1) is reserved for training. This process is repeated until the last fold reserved for testing has been tested. In addition, the whole study was repeated 10 times and averaged and performance values were recorded.  

\begin{figure}[h]
\centering
\includegraphics[width=3.5in,height=1.5in]{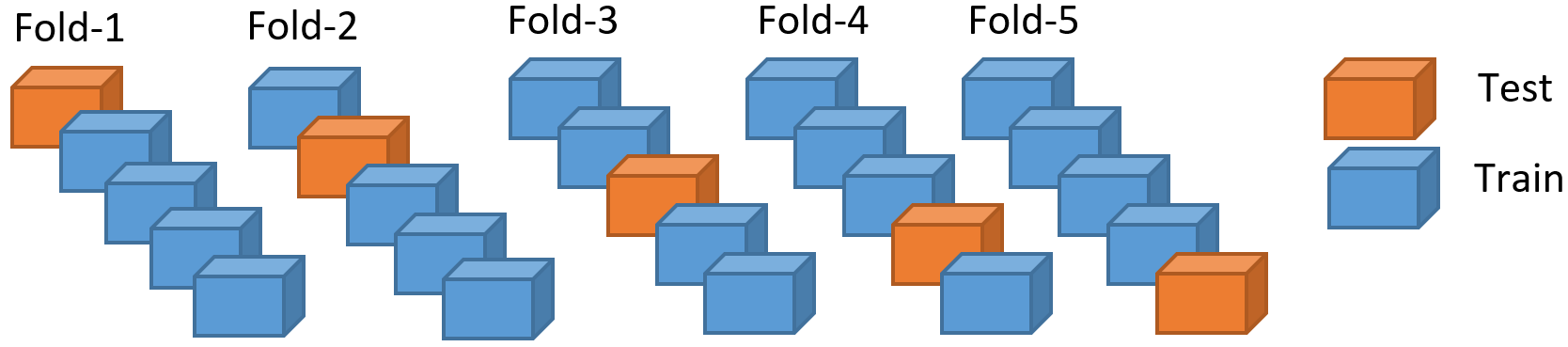}
% where an .eps filename suffix will be assumed under latex, 
% and a .pdf suffix will be assumed for pdflatex; or what has been declared
% via \DeclareGraphicsExtensions.
\caption{Representaton of 5-fold cross validation methods.}
\label{fig_sim}
\end{figure}

\section{RESULTS}
In this study, first of all, all data downloaded from the Kaggle page were made available for the study. In addition, all images were converted from 1024x1024 size to 224x224 size via MATLAB 2020a without pre-processing. All these transformed images were provided as input to the pre-trained ResNet-50 convolutional network model. 1000 features were obtained for each of 2905 data from the layer just before the classification layer. It was submitted to SVM as input with a size of 2905x1000. All these operations performed are given in the flow diagram presented in Figure 3. All the results obtained from this study performed with 3 different functions are presented in Table 1.

\begin{figure}[h]
\centering
\includegraphics[width=3.5in,height=2.0in]{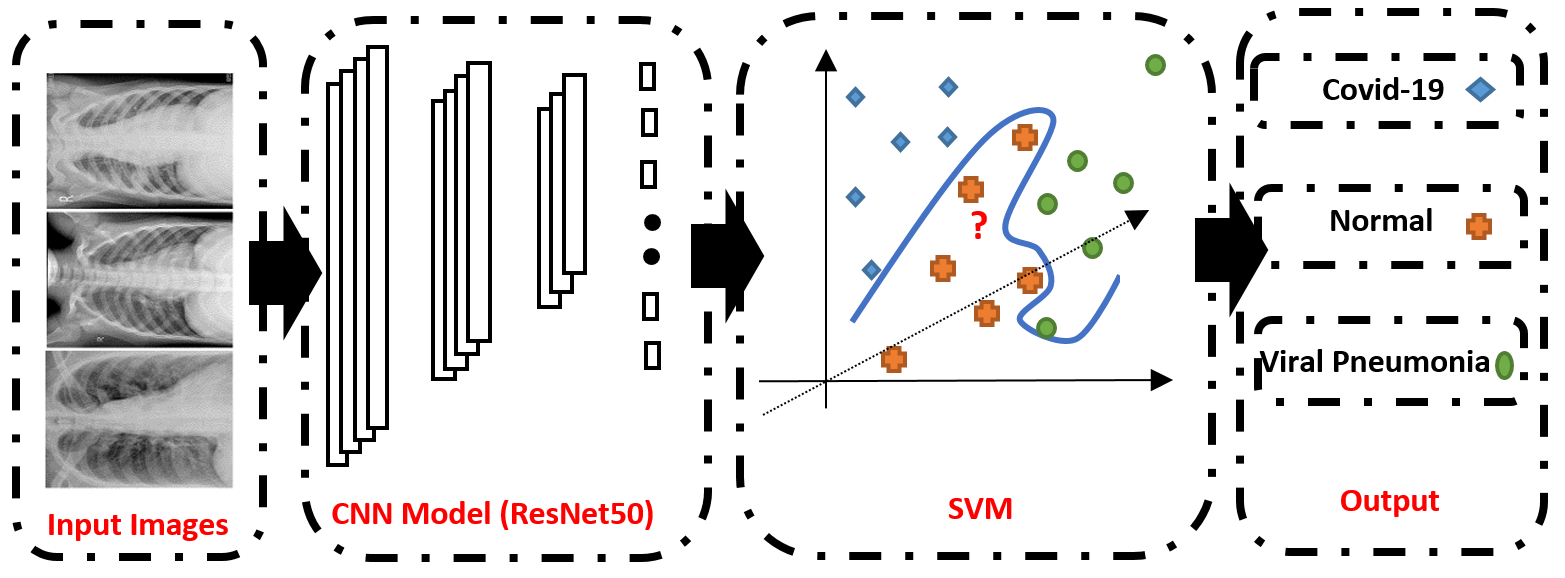}
% where an .eps filename suffix will be assumed under latex, 
% and a .pdf suffix will be assumed for pdflatex; or what has been declared
% via \DeclareGraphicsExtensions.
\caption{Flow diagram of the study.}
\label{fig_sim}
\end{figure}

\begin{table}[h]
\renewcommand{\arraystretch}{2.5}
% if using array.sty, it might be a good idea to tweak the value of
% \extrarowheight as needed to properly center the text within the cells
\caption{\textsc{Performance results for 3 different SVM algorithms.}}
\label{table_example}
\centering
% Some packages, such as MDW tools, offer better commands for making tables
%% than the plain LaTeX2e tabular which is used here.
\begin{tabular}{|l|l|c|c|c|}
\hline
Methods & Class & SEN & SPE & ACC\\
\hline
SVM Linear		&Covid-19&			94.52&	99.70&	98.90\\
				&Normal&			96.87&	95.52&	94.77\\
				&Viral Pneumonia&	94.72&	96.99&	93.65\\
\hline
SVM Quadratic	&Covid-19&			96.35&	99.85&	99.31\\
				&Normal&			97.02&	96.23&	95.28\\
				&Viral Pneumonia&	95.84&	97.37&	94.83\\
\hline
SVM Cubic		&Covid-19&			95.89&	99.96&	99.35\\
				&Normal&			96.57&	95.84&	94.68\\
				&Viral Pneumonia&	95.54&	96.86&	94.30
\\
\hline
\end{tabular}
\end{table}

 The performance of 3 different classes for 3 different SVMs are shown in Table 1. The success of Covid-19 patients detected with the SVM-Quadratic approach is higher than other approaches. At the same time, it is clearly seen that the SVM Quadratic approach gives higher results than other methods in determining healthy individuals, ie normal individuals. For Viral Pneumonia, the SVM-Quadratic approach yielded the highest results. It is clear that the overall performance (ACC) results can also be detected with an accuracy of over 99\%.
In addition to these performance metrics, the Confusion matrixes for all SVM types are given in Figure 4, Figure 5 and Figure 6. It is clearly seen in Figure 5 that 211 of 2019 Covid-19 patients have been successfully detected. The receiver operating characteristic curve (ROC) is also given in  Figure 7 shows that the Area Under the ROC Curve (AUC) value is quite high for the Covid-19 class.

\begin{figure}[h]
\centering
\includegraphics[width=3.5in]{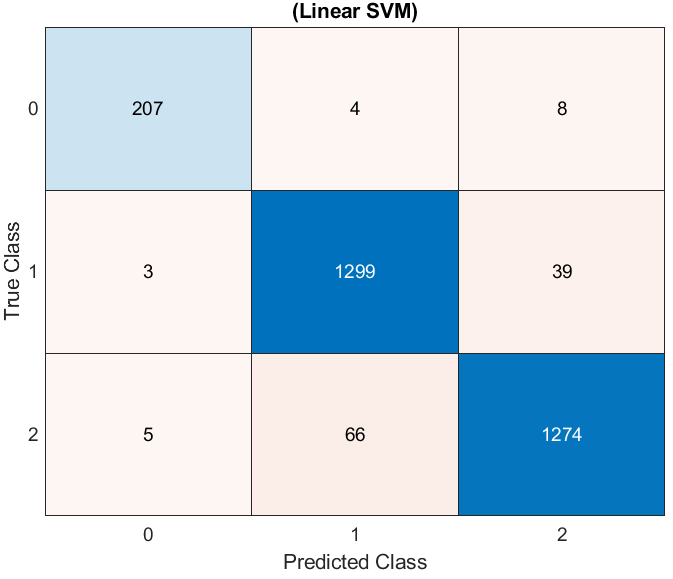}
% where an .eps filename suffix will be assumed under latex, 
% and a .pdf suffix will be assumed for pdflatex; or what has been declared
% via \DeclareGraphicsExtensions.
\caption{Confusion matrix for SVM (linear).}
\label{fig_sim}
\end{figure}

\begin{figure}[h]
\centering
\includegraphics[width=3.5in]{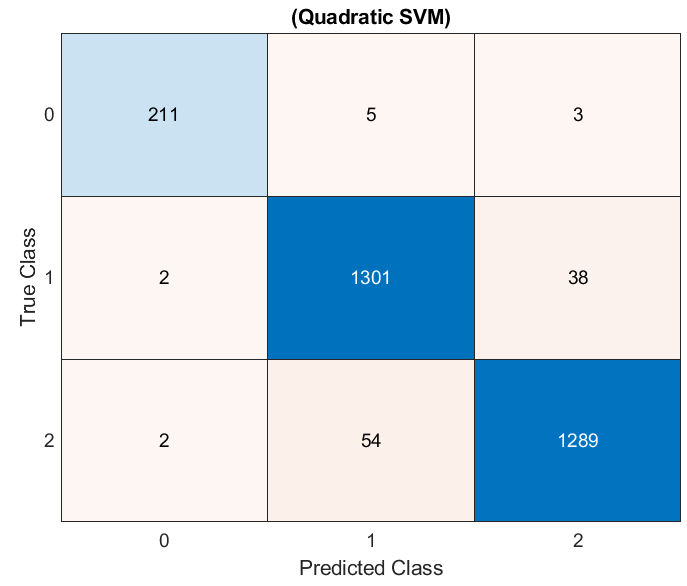}
% where an .eps filename suffix will be assumed under latex, 
% and a .pdf suffix will be assumed for pdflatex; or what has been declared
% via \DeclareGraphicsExtensions.
\caption{Confusion matrix for SVM (quadratic).}
\label{fig_sim}
\end{figure}

\begin{figure}[h]
\centering
\includegraphics[width=3.5in]{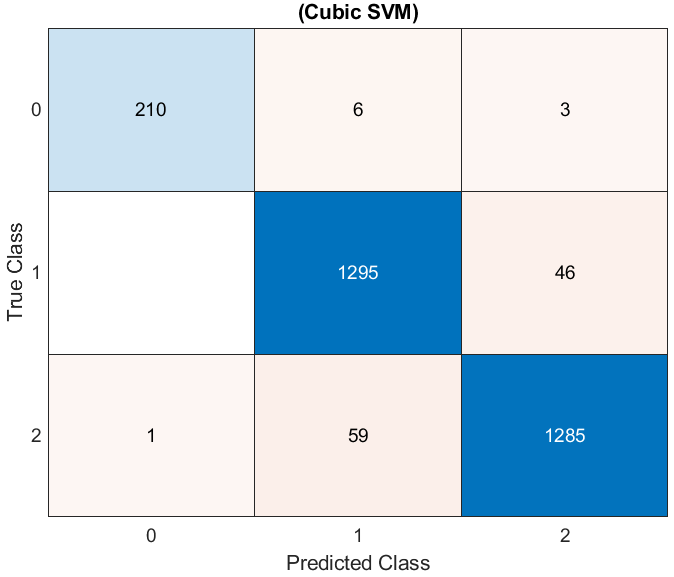}
% where an .eps filename suffix will be assumed under latex, 
% and a .pdf suffix will be assumed for pdflatex; or what has been declared
% via \DeclareGraphicsExtensions.
\caption{Confusion matrix for SVM (cubic).}
\label{fig_sim}
\end{figure}

\begin{figure}[h]
\centering
\includegraphics[width=3.5in]{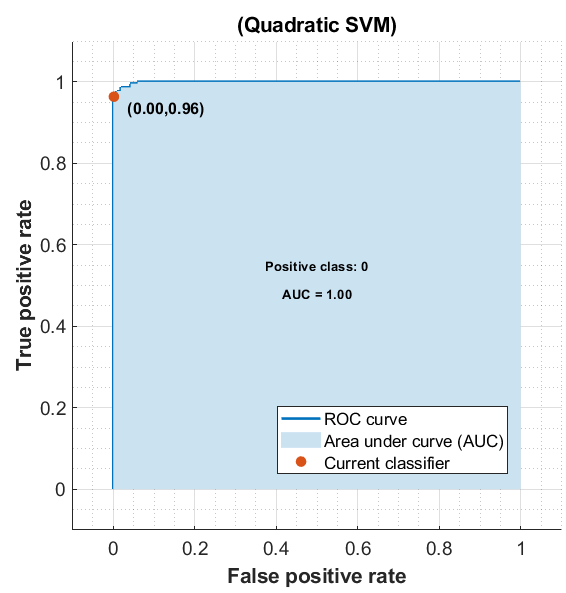}
% where an .eps filename suffix will be assumed under latex, 
% and a .pdf suffix will be assumed for pdflatex; or what has been declared
% via \DeclareGraphicsExtensions.
\caption{ROC curve for SVM (quadratic).}
\label{fig_sim}
\end{figure}

The results obtained are more successful than many studies performed with manual feature extraction methods on Covid-19 in the literature. This study, in which the feature maps obtained from the model are used, is one of the advantageous parts of using the raw data directly and obtaining the features directly without any separate mathematical processing other than the model.

These results are thought to benefit experts during the Covid-19 outbreak. Even if the symptoms are thought to occur late, its use in the follow-up of patients in the most optimistic aspect will relieve the health community.

Based on the results obtained, it is necessary to repeat the study with much more data in order to make a general judgment. In future studies, the data set will be increased and the study will be detailed by using different convolution models.

\section{Conclusion}
In this paper, a multi-class study of Covid-19, Normal and Viral Pneumonia has been made using the SVM algorithm and the features extracted from the X-ray chest images with the help of the ResNet-50 deep learning model. With the study carried out, it has been revealed that the deep learning-based models can be used to extract the features directly and classify them with different classifiers. It gives very high results even for a limited number of Covid-19 data.

\end{document}